\documentclass{aastex63}

\usepackage{graphicx,epsfig}
\usepackage{dcolumn}
\usepackage{bm}
\usepackage{xcolor}
\usepackage{multibib}
\newcites{main}{References}
\newcites{appendix}{References (Appendix)}
\usepackage{subfigure}
\newcommand{\be}{\begin{equation}}
\newcommand{\ee}{\end{equation}}
\newcommand{\bse}{\begin{subequations}}
\newcommand{\ese}{\end{subequations}}
\newcommand{\bary}{\begin{eqnarray}}
\newcommand{\eary}{\end{eqnarray}}
\newcommand{\bwt}{\begin{widetext}}
\newcommand{\ewt}{\end{widetext}}

\submitjournal{APJL}

\begin{document}

\title{Multi-TeV flaring from high energy blazars: An evidence of the
photohadronic process}

\correspondingauthor{Sarira Sahu}

\author{Sarira Sahu}
\email{sarira@nucleares.unam.mx}

\author{Carlos E. L\'opez Fort\'in}
\email{carlos.fortin@correo.nucleares.unam.mx}
\affiliation{Instituto de Ciencias Nucleares, Universidad Nacional Aut\'onoma de M\'exico, \\
Circuito Exterior, C.U., A. Postal 70-543, 04510 Mexico DF, Mexico}

\author{Shigehiro Nagataki}
\email{shigehiro.nagataki@riken.jp}
\affiliation{Astrophysical Big Bang Laboratory, RIKEN,\\
Hirosawa, Wako, Saitama 351-0198, Japan}
\affiliation{Interdisciplinary Theoretical \& Mathematical Science (iTHEMS),\\
RIKEN, Hirosawa, Wako, Saitama 351-0198, Japan}

\begin{abstract}
The high energy peaked blazars are known to undergo episodes of
flaring in GeV-TeV  gamma-rays involving different time
scales and the flaring mechanism is 
not well understood despite long term simultaneous multiwavelength
observations. These gamma-rays {\it en route} to Earth undergo
 attenuation by the extra galactic background light.
Using the photohadronic model, where the seed photons follow a
 power-law spectrum and a template extragalactic
 background light model, we derive a simple relation
 between the observed multi-TeV gamma-ray flux and the intrinsic flux
 with a single parameter.
We study 42 flaring epochs of 23 blazars and excellent fit to
most of the observed spectra are obtained, 
strengthening the photohadronic origin of multi-TeV gamma-rays.
We can also constrain the power spectrum of the seed photons during
the flaring period. The blazars of unknown redshifts, whose multi-TeV flaring spectra are
known, stringent bounds on the former can be placed using the photohadronic model. 
\end{abstract}




\section{Introduction} \label{sec:intro}

Blazars are a subclass of active galactic nuclei (AGNs) which include
flat spectrum radio quasars (FSRQ) and BL Lacertae (BL Lac) objects \citepmain{Romero:2016hjn}.
These objects are characterized by
non-thermal spectra at all wavelengths, from radio to
very high energy (VHE, $> 100$ GeV)  $\gamma$-rays and show flux variability 
on time scales ranging from months to a few minutes \citepmain{Abdo:2010rw}. The flux
variability is produced in a highly relativistic jet pointing towards the observer.
The spectral energy distributions
(SEDs) of blazars are characterized by two non-thermal
peaks \citepmain{Abdo:2009iq}. The first peak (low energy) is located
between infrared to X-ray energies, produced from the
synchrotron emission from the relativistic electrons in the
jet. The general consensus is that the second peak (high energy) 
corresponds to the synchrotron self Compton (SSC) scattering of the
high energy electrons with their self-produced synchrotron photons.
Depending on the location of the first peak, blazars are often
subdivided into low energy peaked blazars (LBLs), intermediate energy
peaked blazars (IBLs) and high energy peaked blazars (HBLs) \citepmain{Abdo:2009iq}.
The leptonic model is
very successful in explaining the multiwavelength emission from
blazars \citepmain{Tavecchio:2010ja,Boettcher:2013wxa}. The 
nearest HBL Markarian 421 (Mrk 421) was the first to be detected in TeV
energy by Whipple telescopes \citepmain{Punch:1992xw}.
In recent years, the highly sensitive Imaging Atmospheric Cherenkov
Telescopes (IACTs) such as VERITAS \citepmain{Holder:2008ux},
HESS \citepmain{Hinton:2004eu} and MAGIC \citepmain{Cortina:2004qt} have great
success in discovering many new extragalactic TeV sources and most of them are
blazars. So, blazars are an important
class of objects to observe and study VHE gamma-ray astronomy.

Flaring in VHE $\gamma$-rays seems to be the major activity of
many HBLs, which is unpredictable and switches between quiescent and
active states involving different time scales \citepmain{Senturk:2013pa}. It
has been 
observed that while in some blazars a strong temporal correlation
between X-ray and multi-TeV $\gamma$-ray exists, in some other, 
except for VHE $\gamma$-rays, no low energy counterpart is observed and explanation
of such anti-correlation is difficult to explain
by leptonic model \citepmain{Krawczynski:2003fq,Blazejowski:2005ih}.
Different models have been developed to explain these flaring
events \citepmain{Giannios:2009pi,Cerruti:2014iwa}. 
Many simultaneous multiwavelength observations have been made to construct
the SED of the flaring period to
constraint different theoretical models \citepmain{Senturk:2013pa,Ahnen:2016hsf}.

The propagating VHE $\gamma$-rays undergo energy dependent
attenuation by the intervening extragalactic background light (EBL)
through pair production \citepmain{Ackermann:2012sza}
and the EBL significantly changes the shape of the VHE spectrum.
So for the calculation of the intrinsic spectrum, a proper understanding
of the EBL SED is important.
Well known EBL models are used by the IACTs collaborations to
analyze the observed VHE $\gamma$-rays from objects of different
redshifts \citepmain{Dominguez:2010bv,Franceschini:2008tp}.

\section*{Photohadronic model}

By assuming that the multi-TeV emission in the HBLs are due to the
photohadronic interaction in the jet \citepmain{Sahu:2019lwj,Sahu:2016mww},
a simple relation 
between the observed VHE spectrum and the
intrinsic spectrum is derived. We
assume that during the VHE emission period, the Fermi accelerated protons
having a power-law spectrum \citepmain{Dermer:1993cz}, $dN/dE\propto
E^{-\alpha}$ (the power index $\alpha \ge 2$) 
, interact with the background seed photons in the jet to produce the
$\Delta$-resonance ($p\gamma\rightarrow \Delta^+$),
which subsequently decays to $\gamma$-rays via intermediate $\pi^0$ 
and to neutrinos through $\pi^+$.
In a canonical jet scenario, the $\Delta$ production efficiency
is very low due to the low photon density. So, to
explain the multi-TeV emission through this process,
super-Eddington power in proton is needed \citepmain{Cao:2014nia}.
To circumvent this problem a double jet structure scenario is proposed \citepmain{Sahu:2019lwj}:
a small compact cone enclosed by a bigger
one along the same axis, and the photohadronic interaction occurs in the
inner jet region.
The photon density $n'_{\gamma,f}$ in the inner compact region
is much higher than the outer region $n'_{\gamma}$ (where prime corresponds to jet comoving
frame) and due to the
adiabatic expansion of the inner jet, its photon density 
decreases by crossing into the outer region. 
As the photon density is
unknown in the inner jet region, we assume a
scaling behavior of the photon densities in the inner and the
outer jet regions, which essentially means that the spectra of
the outer and the inner jets have the same slope. Using this scaling
behavior, we can express the photon density in the inner region in
terms of the photon density of the outer region which is known from
its observed SED.

The kinematical condition to produce
the $\Delta$-resonance is given by \citepmain{Sahu:2019lwj}
\be
E_{\gamma} \epsilon_{\gamma} = 0.032 \, \Gamma{\cal D}(1+z)^{-2}
\mathrm{GeV}^2,
\label{kincond}
\ee
where $E_{\gamma}$, $\epsilon_{\gamma}$,  $\Gamma$, ${\cal D}$ and $z$
are observed VHE $\gamma$-ray, seed photon energy in the observer's frame,
bulk Lorentz factor, Doppler factor and redshift respectively. For
a HBL, $\Gamma\simeq {\cal D}$ is satisfied.
The observed VHE $\gamma$-ray flux depends on the Fermi accelerated
proton flux $F_p$ and the background seed photon density $F_{\gamma,
  obs}(=E^2_{\gamma}\, dN_{\gamma}/dE_{\gamma}) \propto
F_p\, n'_{\gamma,f}$. Also
$F_p\propto E^{-\alpha+2}_\gamma$, and using the scaling
behavior we can express $n'_{\gamma,f}\propto \Phi (\epsilon_{\gamma})
\epsilon^{-1}_{\gamma}$, where $\Phi$ is the observed/fitted flux corresponding to
seed photon energy $\epsilon_{\gamma}$. Previously, the
photohadronic model has been successfully used to explain many flaring HBLs and found that,
for all the cases studied so far, $\Phi$
lies in the tail region of the SSC SED \citepmain{Sahu:2016bdu}.
But this region of the SED is not observed/measured
due to technical difficulties. Mostly leptonic models are used to
calculate the flux in this region and different leptonic models
predict different fluxes. In the same HBL, the flux in this region varies during
different flaring states and also different epochs.
However, irrespective of the model used, the
predicted flux in the tail region of the SSC SED is a power-law given by $\Phi \propto
\epsilon^{\beta}_{\gamma}$ and using the above kinematical condition we
can re-express it as $\Phi \propto E^{-\beta}_{\gamma}$.
Putting everything together and 
taking into account the EBL correction, the observed VHE $\gamma$-ray spectrum
can be expressed as the product of the intrinsic flux $F_{\gamma,int}$
and the attenuation factor due to $e^+e^-$ pair production as,
\be
F_{\gamma,obs}(E_{\gamma})= F_{\gamma,int}(E_{\gamma})
\,  e^{-\tau_{\gamma\gamma}(E_{\gamma},z)} =
  F_0 \, \left (
  \frac{E_{\gamma}}{TeV}  \right )^{-\delta+3} e^{-\tau_{\gamma\gamma}(E_{\gamma},z)},
\label{sed}
\ee
 where, $F_0$ is the normalization constant and
 $\delta=\alpha+\beta$. The
 optical depth $\tau_{\gamma\gamma}$ is a function of $E_{\gamma}$ and $z$.
 $F_0$ and $\delta$ are the only parameters to be
 adjusted to fit the observed spectrum.
 However, strictly speaking the normalization constant is not a free parameter which can be fixed
 from the observed data.
 It is not necessary to know {\it a priori} the value of $\beta$ but
 it can be constrained by fitting the observed data with the parameter $\delta$.
Moreover, the spectral index of the intrinsic differential spectrum
can be defined as $\delta_{int}=-\delta+1$.

The stability of the inner jet on large scales can be estimated from the ratio
$\sigma$ of
the magnetic stress (Poynting flux) and the kinetic stress and for BL Lac objects $\sigma\lesssim 1$.
By considering the generic  values of the parameters, magnetic field $B\sim 1$ G, proton density $n_p\sim
10^{-1}-10^{-2}\, \mathrm{cm^{-3}}$, and bulk Lorentz factor
$\Gamma\sim 10$ we obtain $\sigma\sim 0.4$ which corresponds to a
stable inner jet \citepmain{Cavaliere:2017cwv}.
The photon density within the inner jet region can be constrained by
comparing the jet expansion timescale $t'_d$ with the $p\gamma$ interaction
timescale $t'_{p\gamma}$ and
assuming the high energy proton luminosity to be smaller than the
Eddington luminosity \citepmain{Sahu:2015tua}. 

\section*{Results and analysis}
Using Eq. (\ref{sed}) we fitted the observed VHE spectra of 42
emission epochs of 23 HBLs of different redshifts very well with
 the free parameter $\delta$ is in the range 
$2.5\le\delta \le 3.0$.
Depending on the value of $\delta$, we roughly classify these flaring
 states into three different categories as follows:
  (i) low state, when $\delta = 3.0$, (ii) high state, when $2.6< \delta < 3.0$,
  and (iii) very high state, when $2.5 \le \delta \le 2.6$.
We know {\it a priori} that $\alpha \ge 2$, so during the simultaneous observation
period in multiwavelength, we must have $ 0.0\le
\beta \le 1.0$.
The three different emission states are discussed through four
examples with HBLs of different redshifts and the
EBL model of Fransceschini et
al \citepmain{Franceschini:2008tp} is used for our analysis.

\subsection*{1ES 0229+200}
\begin{figure}
\centering
\includegraphics[width=0.9\textwidth]{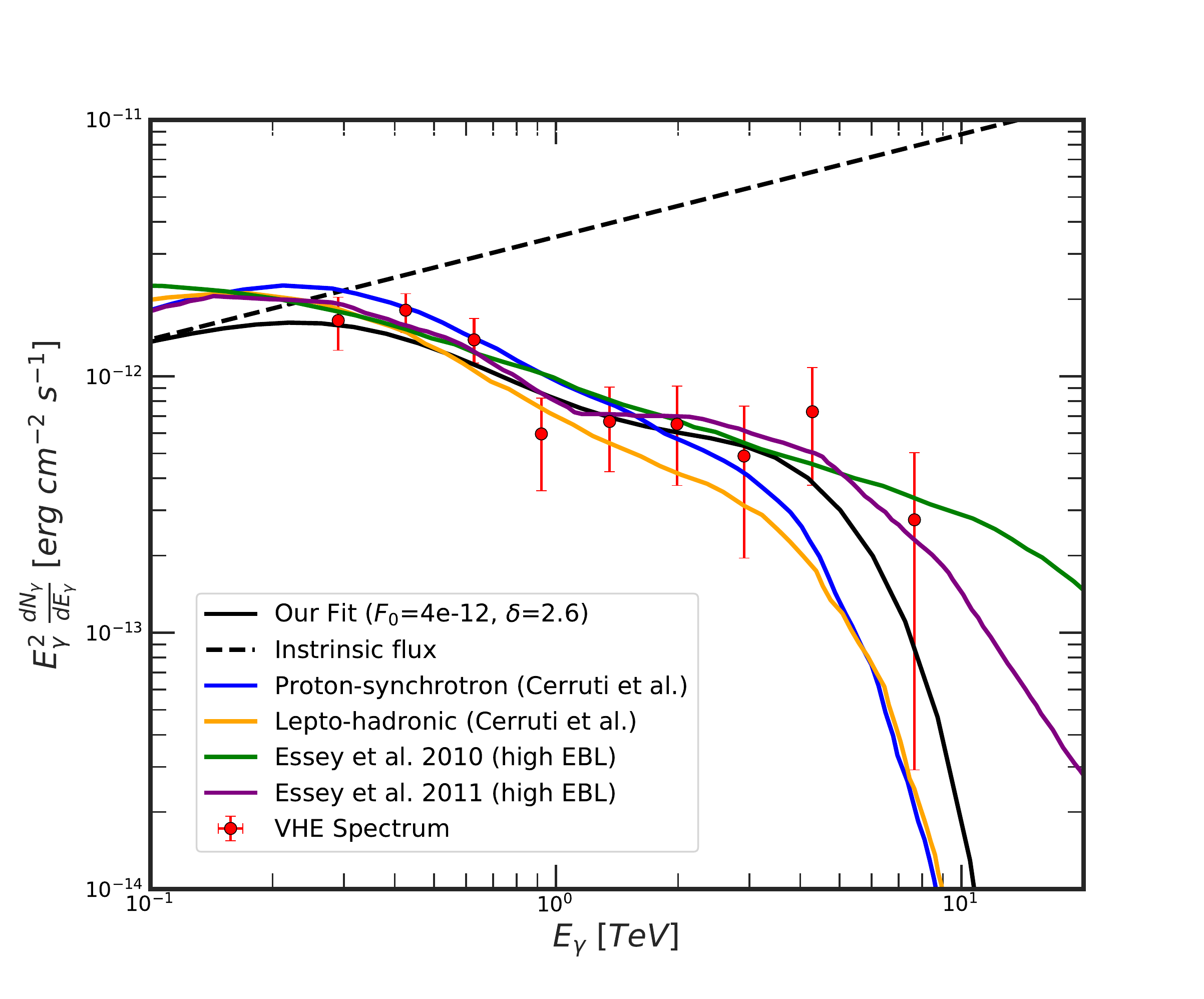}
\caption{\small \textbf{Multi-TeV SED of 1ES 0229+200}.
The time-averaged observed spectrum (red data points) of HBL 1ES
0229+200 during October 2009 and 
January 2013 by VERITAS telescopes \citepmain{Aliu:2013pya} is shown.
An excellent fit is obtained with the photohadronic model with
$\delta=2.6$ and $F_0=4.0\times 10^{-12}
\mathrm{erg\, cm^{-2}\, s^{-1}}$ (black curve) and the corresponding intrinsic flux is also shown 
(black dashed curve). In all the subsequent figures
the values of $\delta$ and $F_0$ (in $\mathrm{erg\, cm^{-2}\,
  s^{-1}}$ unit) are given in the legend.
For comparison we have
also shown the proton-synchrotron fit and the lepto-hadronic
fit \citepmain{Cerruti:2014iwa} and the hadronic model \citepmain{Essey:2009ju,Essey:2010er}.
\label{fig:fig1}
}
\end{figure}

The 1ES 0229+200 is a HBL at a redshift of $z=0.1396$ which was
discovered in the Einstein IPC Slew Survey in 1992 \citepmain{Schachter:1993apj}.
It was observed by VERITAS telescopes during
a long-term observation over three seasons between October 2009 and
January 2013, for a total of 54.3 hours \citepmain{Aliu:2013pya} and an excess
of 489 $\gamma$-ray events were detected in the energy range
$0.29 \mathrm{TeV} \le E_\gamma \le 7.6 \mathrm{TeV}$.
Using the proton-synchrotron model and
the lepto-hadronic model, dominated by emission from the secondary
particles from $p\gamma$ interactions, the observed multiwavelength
SEDs of several HBLs are fitted by Cerruti {\it et al.} \citepmain{Cerruti:2014iwa}. However, these numerical models use about 19 parameters to
fit the entire SED. We have shown their fit to 1ES 0229+200 in
Figure \ref{fig:fig1}. Alternatively, using the photohadronic model an excellent fit is obtained
for $\delta=2.6$ and $F_0 = 3.5\times 10^{-12}\,
\mathrm{erg\,cm^{-2}\,s^{-1}}$. According to the above discussed
classification scheme, this corresponds to very high emission state
and the intrinsic flux $F_{int}\propto
E^{0.4}_{\gamma}$. Similarly, the extracted differential spectrum
$(dN_{\gamma}/dE_{\gamma})_{int}\propto E^{-1.6}_{\gamma}$ which is not
hard. 
This HBL has the central black hole of mass $M_{BH}\sim 1.4\times 10^9 M_{\odot}$
and outer blob size $R'_{b}\sim 10^{16}-10^{17}$ cm \citepmain{Zacharopoulou:2011yr}. Assuming the
high-energy proton flux corresponding to $E_\gamma=7.6$ TeV to be smaller than the Eddington flux and
comparing $t'_d$ (inner blob size $R'_{f}\sim 4\times10^{15}$ cm) with
$t'_{p\gamma}$, we obtain the photon density in the range $4\times
10^8\, \mathrm{cm^{-3}}<n'_{\gamma,f}<2.5\times
10^{11}\, \mathrm{cm^{-3}}$.

From this HBL between 2005 and 2006,  the HESS telescopes also observed VHE $\gamma$-rays \citepmain{Aharonian:2007wc} whose time-averaged spectrum is in the
energy range $0.5\, \mathrm{TeV} \le E_\gamma \le 11.5\,
\mathrm{TeV}$ and is very similar to the one discussed
above. This spectrum is fitted with the hadronic model of Essey {\it et al.}
\citepmain{Essey:2009ju,Essey:2010er}. 
Using the photohadronic model a very good fit is obtained for
$\delta=2.5$ (see Figure 7 of Supplementary Materials for details).
By reducing 10\% to the hadronic model of Essey {\it et al.} the
spectrum of VERITAS can be fitted well, which is shown in Figure \ref{fig:fig1} for comparison.

\subsection*{1ES 0347-121}
The 1ES 0347-121 is a HBL at a redshift of $z=0.188$. The HESS telescopes observed
this blazar between August and December 2006 for a total of 25.4
hours \citepmain{Aharonian:2007tc} when an excess of 327 VHE gamma-ray events
were detected in the energy range $0.25\, \mathrm {TeV} \le E_\gamma \le 3$ TeV
and no flux variability was detected in the data set.

\begin{figure}
\centering
\includegraphics[width=0.9\textwidth]{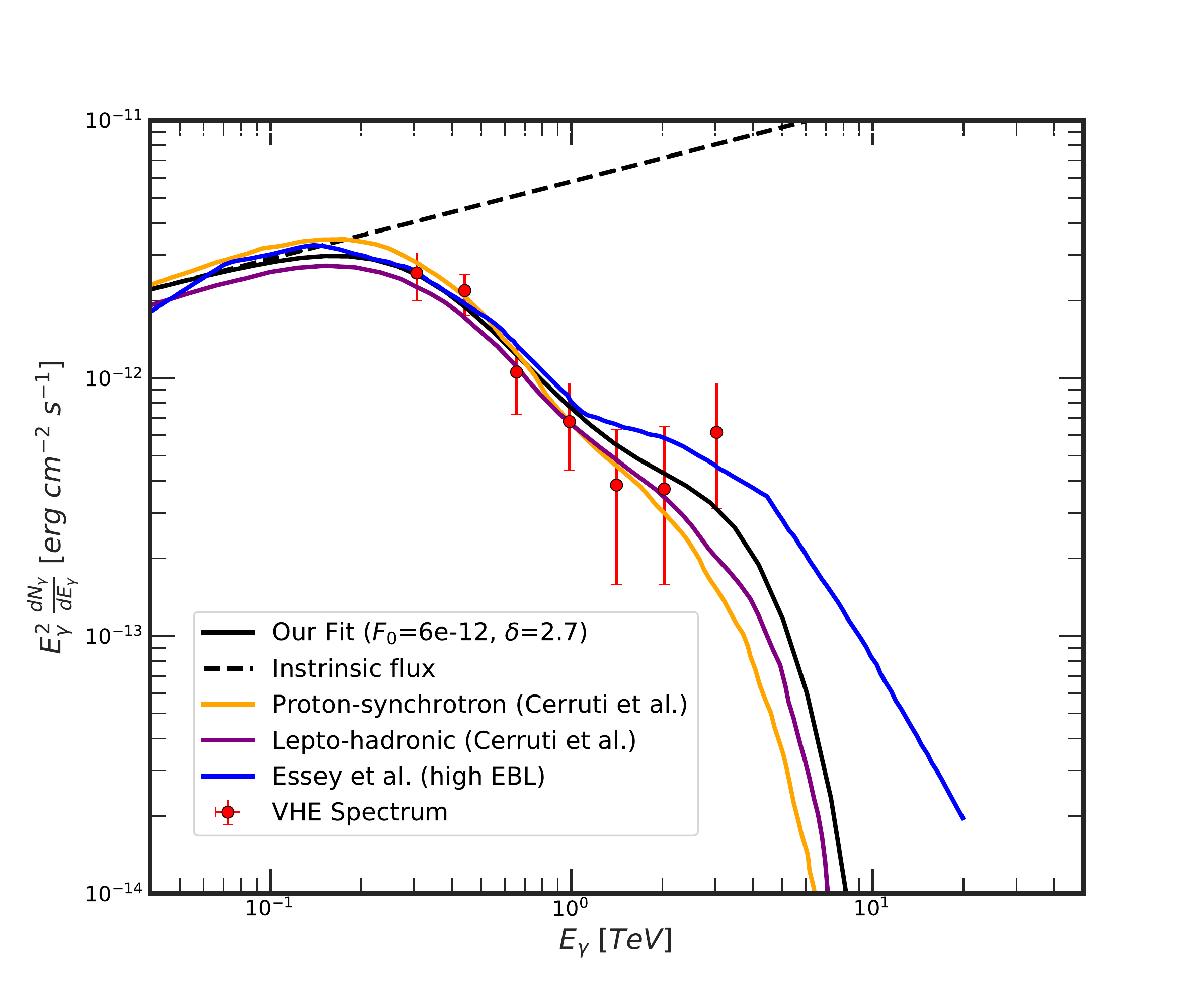}
\caption{\small \textbf{Multi-TeV SED of 1ES 0347-121}.
The VHE spectrum of HBL 1ES 0347-121 observed by the HESS telescopes between
August and December 2006 \citepmain{Aharonian:2007tc} is fitted using
photohadronic model (black
curve) and its corresponding intrinsic spectrum is shown (black
dashed curve). Our
result is compared with the hadronic model of 
Essey et al. (high EBL) \citepmain{Essey:2010er} (blue curve) and the proton-synchrotron
model (blue curve) and lepto-hadronic model (orange curve) \citepmain{Cerruti:2014iwa}.
\label{fig:fig2}
}
\end{figure}

In a hadronic model scenario, ultra high energy protons escaping from the jet
produce secondary VHE gamma-rays by interacting with the Cosmic
Microwave Background (CMB) and/or EBL \citepmain{Essey:2010er}. Assuming this
scenario the spectra of 1ES 0347-121, 1ES 0229+200 and
1ES 1101-232 are explained well \citepmain{Essey:2010er}. However,
this scenario requires protons in the energy range $10^{8}$–$10^{10}$
GeV which are not easily produced in the jet environment, as well as a
weak extragalactic magnetic field in the range
$10^{-17}\, \mathrm {G}< \mathrm {B}< 10^{-14}$ to produce the observed
gamma-ray spectrum along the line of sight \citepmain{Essey:2010nd}. In an alternative
scenario, 
Cerruti {\it et al.} \citepmain{Cerruti:2014iwa}
have applied the proton-synchrotron and lepto-hadronic models to fit
the spectrum of 1ES 0347-121.
Using the photohadronic model we found an excellent fit to the
spectrum with $\delta=2.7$
and $F_0=6.0\times 10^{-12}\, \mathrm {erg\,cm^{-2}\,s^{-1}}$, which is
a high state emission.
As $\gamma$-ray carries $\sim 10\%$ of the proton
energy \citepmain{Sahu:2019lwj}, $E_{\gamma}=3$ TeV corresponds to 30 TeV
cosmic ray proton energy which can easily be produced and accelerated
in the blazar jet. In  Figure \ref{fig:fig2} we compare our result with
Essey {\it et al.} \citepmain{Essey:2010er} and Cerruti et
al.  \citepmain{Cerruti:2014iwa} and found that below 1 TeV all have similar
behaviors. However, above 1 TeV our result differs substantially from
the others, particularly from Essey {\it et al.} which uses the EBL
model of Stecker {\it et al.} (high EBL) \citepmain{Stecker:2006aj}.
Comparison of the EBL models of
Franceschini {\it et al.} \citepmain{Franceschini:2008tp} and Stecker {\it et al.} \citepmain{Stecker:2006aj}
shows a significant difference in the attenuation factor above 1 TeV. 

\subsection*{1ES 0806+524}
\begin{figure}
\centering
\includegraphics[width=0.9\textwidth]{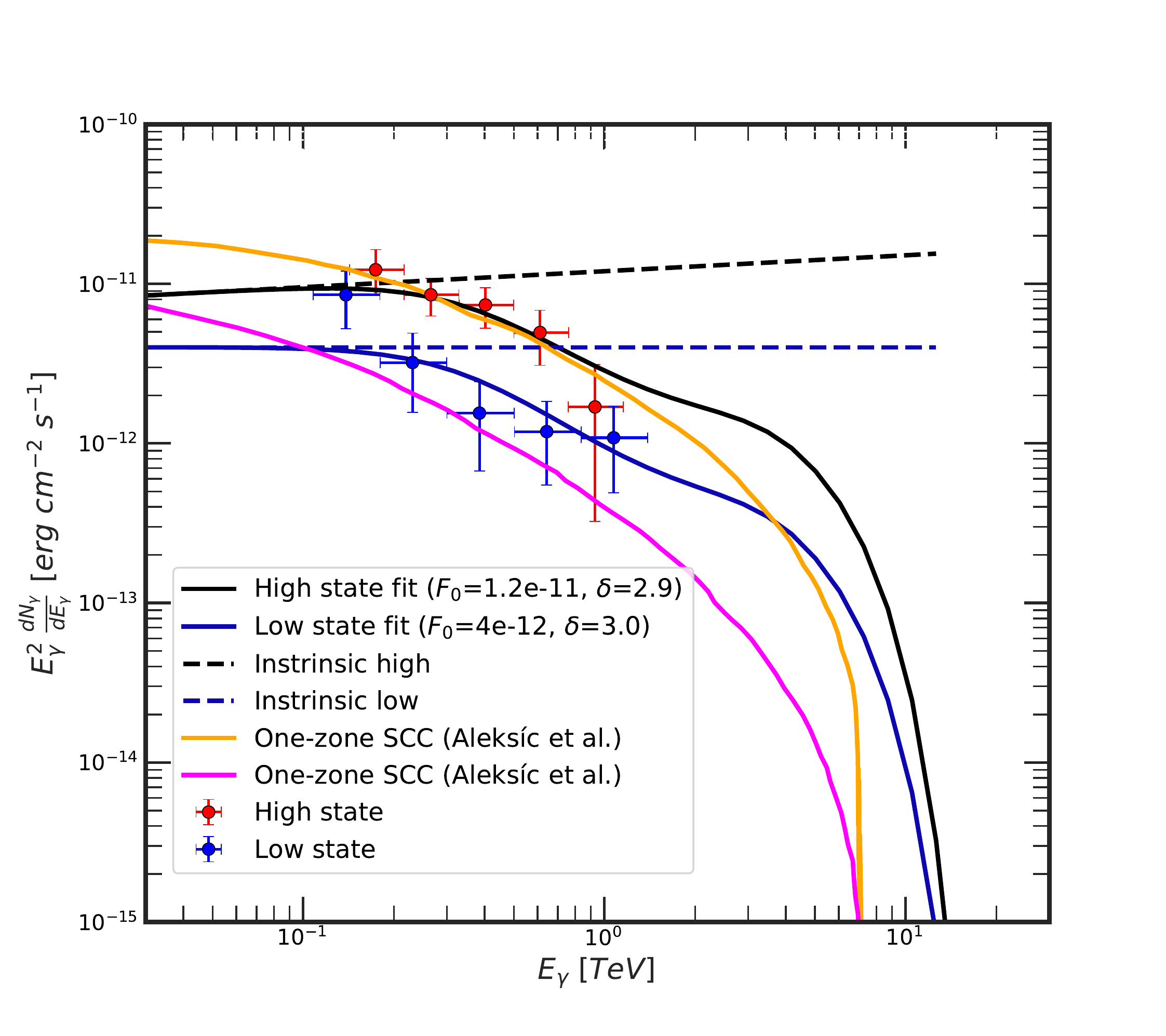}
\caption{\small \textbf{Multi-TeV SED of 1ES 0806+524}.
The MAGIC observation of the HBL 1ES 0806+524 from January
to March 2011 is shown here. A flaring event was observed on 24
February. The observed fluxes for both the flaring (red data points) and
the average of the remaining data (blue data points) are shown. They are fitted using
one-zone SSC model \citepmain{Aleksic:2015ula} and the photohadronic model (black curve).
\label{fig:fig3}
}
\end{figure}

The 1ES 0806+524 is at a redshift of $z=0.138$ and in 2008, the VERITAS
telescopes discovered this in VHE
$\gamma$-rays \citepmain{Acciari:2008cn}.
A multiwavelength observation was performed by MAGIC telescopes from January to
March 2011 for 13 nights for about 24
hours \citepmain{Aleksic:2015ula} and, on February 24, observed a flaring event. Within 3
hours of observation excess events above 250 GeV were recorded
in the energy range $0.17\, \mathrm {TeV} \le E_\gamma \le 0.93\,
TeV$ when, the flux increased by a factor of about 3 from the mean
flux level and no intra-night variability was observed.
The flaring data and the remaining MAGIC
observations are analyzed separately using the photohadronic
scenario, which are shown in  Figure \ref{fig:fig3}. Using one-zone SSC
model, the broad-band SEDs during the flaring (high) and the remaining
period (low) are explained using about 14 free
parameters. The electron Lorentz factor for the high state is double
the one for the low state and the remaining parameters are the same \citepmain{Aleksic:2015ula}.
With the photohadronic scenario, the flaring
state can be fitted very well with $\delta=2.9$ and $F_0=1.2\times
10^{-11}\, \mathrm{erg\,cm^{-2}\,s^{-1}}$ corresponding to a high
emission state and the average of the remaining flux can be
fitted with $\delta=3.0$ and $F_0=4.0\times
10^{-12}\, \mathrm{erg\,cm^{-2}\,s^{-1}}$, which is a low state. The intrinsic
fluxes in high and low states are respectively proportional to
$E^{0.1}_\gamma$ and $E^0_\gamma$. Comparison of both the models is
shown in the Figure \ref{fig:fig3}. The SSC model does not fit well to
the low state spectrum. Although both models explain well the
flaring data, a significant difference is observed in the
predictions above 1 TeV.

\subsection*{HESS J1943+213}

\begin{figure}
\centering
\includegraphics[width=0.9\linewidth]{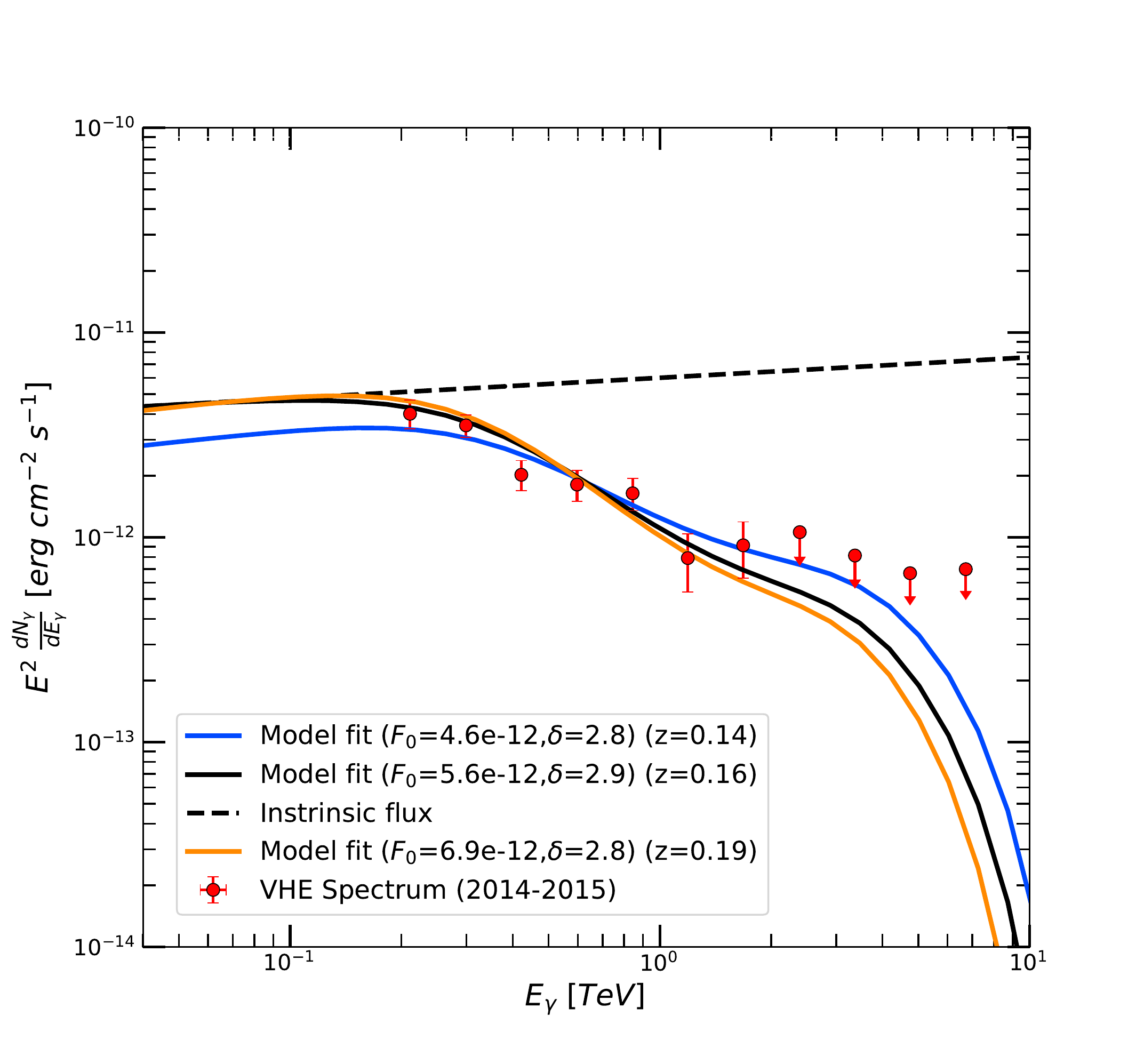}
\caption{\small \textbf{Multi-TeV SED of HESS J1943+213}.
The EHBL HESS J1943+213 has unknown redshift and it was
observed in VHE by VERITAS from 27 May 
to 2 July 2014, and 20 April to 9 November 
2015. The time-averaged spectrum of both
observations is shown \citepmain{Archer:2018ehn}. Using the photohadronic model
and performing a statistical analysis for
different redshifts, we were able to constrain the 
redshift in the range $0.14\le z\le 0.19$. The values of $\delta$ and
$F_0$ are also shown in the figure.
}
\label{fig:fig4}
\end{figure}

HESS J1943+213 is a VHE gamma-ray point source discovered by
HESS \citepmain{Abramowski:2011ri} which is identified as an extreme HBL (EHBL).
In VHE, it was observed by VERITAS telescopes from 27 May
to 2 July 2014 and from 20 April to 9 November 2015, a total exposure time of 37.2
hours and no flux variability was observed \citepmain{Archer:2018ehn}.  The
time-averaged spectrum of both the observations is presented in Figure \ref{fig:fig4}.
Currently, the redshift of HESS J1943+213 is not known and indirect
limits ($0.03< z < 0.45$) were set by Peter {\it et al} \citepmain{Peter:2014hga}. 
Improved gamma-ray spectra of Fermi-LAT and VERITAS were used to derive a
conservative upper limit of $z < 0.23$ \citepmain{Archer:2018ehn}. Using the photohadronic model
and different redshifts, we derived more stringent lower and
upper limits on the redshift ($0.14\le z\le 0.19$) which are shown in
Figure \ref{fig:fig4}. However, the best fit is obtained for $z=0.16$
and $\delta=2.9$ corresponding to a high state emission from the
source. Additional two such examples are discussed in the
supplementary materials. 


\section*{Discussion}

The HBLs are known to undergo episodes of VHE flaring in
gamma-rays involving
different time scales and the flaring mechanism is not well
understood. Also the VHE gamma-rays are attenuated by EBL background.
Here we have derived a simple relation between the observed VHE flux
and the intrinsic flux from the flaring HBLs by assuming that
during flaring, Fermi-accelerated high energy protons
interact with the seed photons in the inner compact region of the
jet to produce $\Delta$-resonance which subsequently decays to
gamma-rays and neutrinos from intermediate $\pi^0$ and $\pi^+$
respectively. These gamma-rays can be observed.
To account for the EBL effect we consider the well known EBL model of
Franchesccini {\it et al}.
and analyzed 23 HBLs of different redshifts and a total of 42 different emission
epochs of them. For detailed analysis we only used five emission
epochs of four HBLs, and the rest of the flaring states are 
summarized in Table 1. Some of these are briefly discussed in
supplementary materials to strengthen further the validity of the
photohadronic origin of multi-TeV flaring events.

From the analysis we observed that the free 
parameter $\delta$ is constrained to be in the range $2.5\le
\delta\le3.0$.
The intrinsic flux for the low state is a
constant, but for high and very high state it is a power-law
proportional to $E^{\eta}_{\gamma}$, where $0 <\eta \le 0.5$.
We could not find any flaring state which has $\delta < 2.5 $.
Some flaring spectra can be fitted well with $\delta > 3$. 
However, it is important to note that for these
cases $-\delta+3$ is positive (a very soft
spectrum) and in the low-energy limit the spectrum shoots up very high,
which is certainly not observed. So the soft power-law fits are
 ignored \citepmain{Dwek:2004pp,Sahu:2018gik} and we always adhere to $\delta \le 3.0$.
From the analysis we
observed that about 48\% are low states, 38\% high states and 14\%
are very high state emissions. This implies that low and high emission
states constitute the major part of the flaring in HBLs.
 
Although, the photohadronic scenario works well for $E_{\gamma}
\gtrsim 100$ GeV, there are contributions from the leptonic processes
 to the observed spectrum in this energy regime, so in the
low energy regime our model may not fit the data very well. In some
cases, we have
observed that the averaging of long-term VHE observations are difficult
to explain by photohadronic model for the following reasons:
gamma-rays from the leptonic processes contribute to the spectrum in the
low energy regime and the averaging of many unobserved short flares with
the low emission periods contaminates the data.

Several models explain well the observed broadband SEDs but
require many assumptions and many free parameters, some of which are difficult to realize
in the jet environment \citepmain{Cerruti:2014iwa,Essey:2010er,Aleksic:2015ula,Boettcher:2013wxa}. On the other hand, the photohadronic scenario
is based on very simple assumptions which are 
very likely to be realized in the jet during the VHE emission
period. Another important aspect of our model is that, the assumption
of the power-law behavior of the 
background seed photon is sufficient to fit the observed spectrum and
it is not necessary to have simultaneous multiwavelength observations.
Moreover, the exact simultaneous
multiwavelength observation during a flaring event in a HBL is usually limited
to a few. In our case, the IACTs observations are sufficient.
From the fitting to the observed spectrum and using 
$\alpha\ge 2$, the seed photon 
spectral index $\beta$ can be constrained. For example, an excellent
fit to the flaring of PG 1553+113 is obtained for $\delta=2.5$
which shrinks the $\beta$ value in the interval 0 to 0.5 (see Table 1).
Nevertheless, the fact that we can explain very well the VHE spectra of
42 epochs of 23 HBLs with a single parameter, provides strong evidence
that VHE gamma-rays are produced mostly through
the photohadronic process with the intermediate $\Delta$-resonance. 
In addition, it is important to mention that for HBLs of unknown
redshifts, whose multi-TeV spectra are known, stringent
bounds on the redshifts can be placed using the photohadronic model.

\acknowledgments

The work of S.S. is partially supported by
  DGAPA-UNAM (Mexico) Project No. IN103019. S.N. is partially
  supported by ``JSPS Grants-in-Aid for Scientific Research $<$KAKENHI$>$
  (A) 19H00693'', 
``Pioneering Program of RIKEN for Evolution of Matter in the Universe
(r-EMU)'', and 
``Interdisciplinary Theoretical and Mathematical Sciences Program of RIKEN''.

\clearpage

\clearpage

\clearpage

\clearpage

\clearpage
\begin{table}
\label{table.1}
\footnotesize
\begin{center}
\begin{tabular}{cccccc}
Name & Redshift($z$) & Period & $F_{0,11}$ & $\delta$ & State \\ [1 mm]
\hline 
Mrk 421 & 0.031 & 2004 & $51.3$ & 2.95 & High \\ 
\ & \ & 22 Apr 2006 & $5.2$ & 2.95 & High \\
\ & \ & 24 Apr 2006 & $10.7$ & 3.0 & Low \\ 
\ & \ & 25 Apr 2006 & $6.9$ & 2.95 & High \\ 
\ & \ & 26 Apr 2006 & $5.2$ & 3.0 & Low \\ 
\ & \ & 27 Apr 2006 & $16$ & 2.95 & High \\ 
\ & \ & 28 Apr 2006 & $5.0$ & 3.0 & Low \\
\ & \ & 29 Apr 2006 & $4.9$ & 3.0 & Low \\ 
\ & \ & 30 Apr 2006 & $13.5$ & 2.5 & Very High \\ 
\ & \ & 16 Feb 2010 & $12$ & 3.0 & Low \\
\ & \ & 17 Feb 2010 & $1.5$ & 3.0 & Low \\
\ & \ & 10 Mar 2010 & $21$ & 2.6 & Very High \\
\ & \ & 10 Mar 2010 & $16.5$ & 3.0 &  Low \\ 
\ & \ & 28 Dec 2010 & $6.7$ & 3.00 & Low \\ 

Mrk 501 & 0.034 & 22 - 27 May 2012 & $6.3$ & 2.9 & High \\
\ & \ & 23 - 24 Jun 2014 & 28 & 2.93 & High \\

1ES 2344+514 & 0.044 & 4 Oct 2007 - 11 Jan 2008 & $0.8$ & 3.0 & Low \\ 

1ES 1959+650 & 0.048 & May 2002 & $12$ & 3.0 & Low \\ 
\ & \ & Nov 2007 - Oct 2013 & $2.2$ & 3.0 & Low \\
\ & \ & 21-27 May 2006 & $1.1$ & 3.0 & Low \\ 
\ & \ & 20 May 2012 & $80$ & 2.9 & High \\

1ES 1727+502 & 0.055 & 1-7 May 2013 & $0.9$ & 3.0 & Low \\ 

PKS 1440-389 & 0.14$\le$$z$$\le$0.24 & 29 Feb - 27 May 2012 & $0.90$ & 3.0 & Low \\

1ES 1312-423 & 0.105 & Apr 2004 - Jul 2010 & $0.20$ & 3.0 & Low \\

B32247+381 & 0.119 & 30 Sep - 30 Oct 2010 & $0.17$ & 3.0 & Low \\ 

RGB J0710+591 & 0.125 & Dec 2008 - Mar 2009 & $0.5$ & 2.9 & High \\ 

1ES 1215+303 & 0.131 & Jan - Feb 2011 & $90$ & 3.0 & Low \\ 

1RXS J101015.9-311909 & 0.14 & Aug 2008 - Jan 2011 & $0.2$ & 2.8 & High \\ 

1ES 0229+200 & 0.14 & 2005 - 2006 & $0.4$ & 2.5 & Very High \\

H 2356-309 & 0.165 & Jun - Dec 2004 & 0.3 & 2.9 & High \\

1ES 1218+304 & 0.182 & Dec 2008 - 2013 & $1.5$ & 2.9 & High \\

1ES 1101+232 & 0.186 & 2004 - 2005 & $0.60$ & 2.75 & High \\ 

1ES 1011+496 & 0.212 & 6 Feb - 7 Mar 2014 & $8.2$ & 3.0 & Low \\ 

1ES 0414+009  & 0.287 & Aug 2008 - Feb 2011 & $0.70$ & 2.9 & High \\ 

PG 1553+113 & 0.50 & 26 - 27 Apr 2012 & $48$ & 2.5 & Very High \\ 

RGB J0152+017 & 0.80 & 30 Oct - 14 Nov 2007 & $0.3$ & 3.0 & Low \\ 

RGB J2243+203 & 0.75$\le$$z$$\le$1.1 & 21 - 24 Dec 2014 & $0.28$ & 2.6
                                                      & Very High \\ [1 mm]
 
\hline
\end{tabular}
\caption{\small Flaring states of the additional HBLs (besides the
  ones already discussed) are shown here. In the 4th column the
  normalization factor is expressed in units of $F_{0,11}=1.0\times\,10^{-11}\, \mathrm {erg\,cm^{-2}\,s^{-1}}$.
 The photohadronic fits to some of these emission states are included in the Supplementary materials.}
\end{center}
\end{table}

\bibliographystylemain{aasjournal}
\bibliographymain{Paper_v1_3}


\end{document}


\title{Supplementary Materials for:\\ Multi-TeV flaring from high energy blazars: An evidence of the
  photohadronic process}

\author{Sarira Sahu}
\email{sarira@nucleares.unam.mx}

\author{Carlos E. L\'opez Fort\'in}
\email{carlos.fortin@correo.nucleares.unam.mx}
\affiliation{Instituto de Ciencias Nucleares, Universidad Nacional Aut\'onoma de M\'exico, \\
  Circuito Exterior, C.U., A. Postal 70-543, 04510 Mexico DF, Mexico}

\author{Shigehiro Nagataki}
\email{shigehiro.nagataki@riken.jp}
\affiliation{Astrophysical Big Bang Laboratory, RIKEN,\\
  Hirosawa, Wako, Saitama 351-0198, Japan}
\affiliation{Interdisciplinary Theoretical \& Mathematical Science (iTHEMS),\\
  RIKEN, Hirosawa, Wako, Saitama 351-0198, Japan}

\renewcommand{\thetable}{\arabic{table}}
\renewcommand{\thefigure}{\arabic{figure}}
\setcounter{figure}{4} 
\setcounter{table}{1} 
\section*{Supplementary Figures}\label{sup:intro}

Due to space constraints in the main article, we only
analyzed five flaring states of four HBLs in the context of the
photohadronic model and compared it with other available models.
However, to further support the validity of our model and its predictions, here, we
provide eleven additional flaring states of HBLs of different
redshifts. Particularly, our best fits to the flaring events of 1ES 0229+200 and
1ES 1101+232 are compared with other existing leptonic and hadronic models,
where we observed that our results are as good as or better than
these models. The redshifts of the HBLs PKS 1440-389 and RGB J2243+203 are
unknown and, using different observations, limits were set to the
redshifts. We have shown that the predicted photohadronic model limits are
more stringent than the existing ones. The references to all the
additional HBLs given in Table 1 are shown in  Table 2.

\begin{figure}
  \centering
  \includegraphics[width=0.9\linewidth]{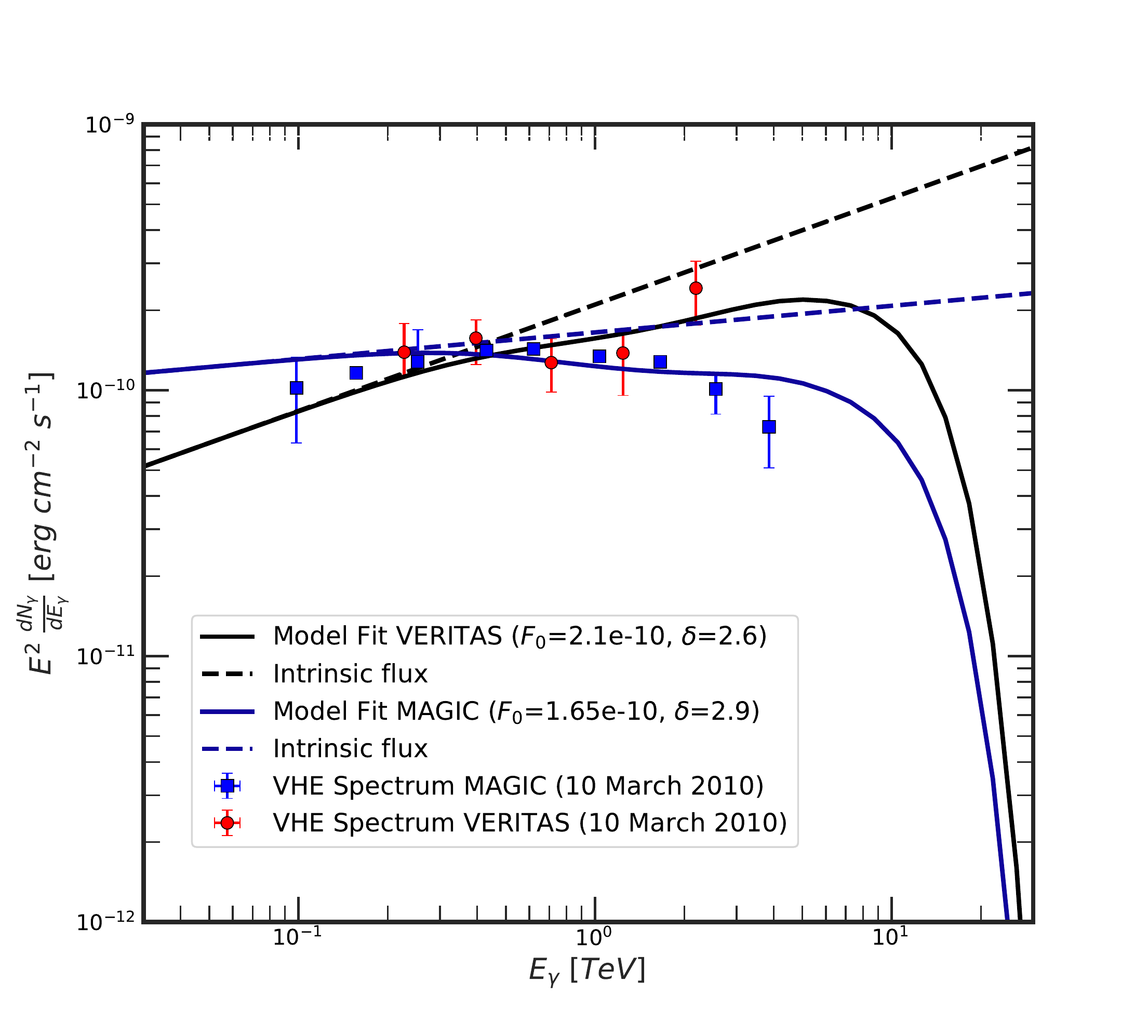}
  \caption{\small \textbf{Multi-TeV SED of Mrk 421}.
    During a multiwavelength campaign of Mrk 421 in March 2010, an ongoing
    VHE flare was observed for 13 consecutive days from 10 to 22
    March \citep{Aleksic:2014rca}. Initially the flare was high and slowly
    decreased during the 13-day period, which was observed by both MAGIC
    and VERITAS telescopes. VERITAS observed high VHE flux on 10 March
    which is roughly 50\% higher than the flux measured by MAGIC for that
    same day. Using the photohadronic model we fitted well with
    $F_0=1.65\times 10^{-10} \, \rm{erg\, cm^{-2}\, s^{-1}}$,
    $\delta=2.9$ for the MAGIC spectrum, which is high, and $F_0=2.1\times
    10^{-10} \, \rm{erg\, cm^{-2}\, s^{-1}}$, $\delta=2.6$ for the
    VERITAS spectrum, which is very high. The corresponding intrinsic
    spectra are shown in dashed lines.}
  \label{fig:sfig1}
\end{figure}

\clearpage

\begin{figure}
  \centering
  \includegraphics[width=\linewidth]{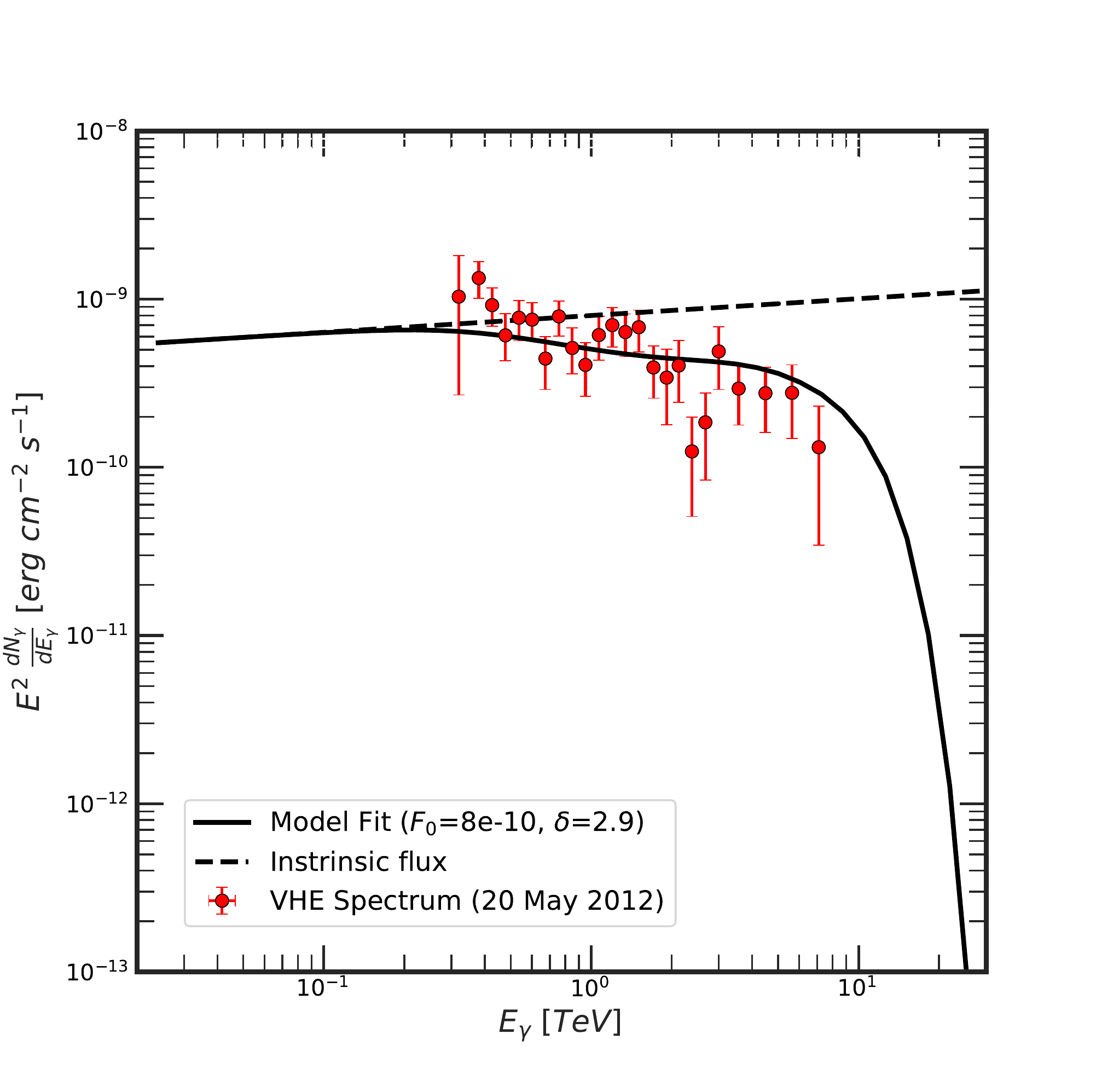}
  \caption{\small \textbf{Multi-TeV SED of 1ES 1959+650}.
    The VERITAS telescopes observed VHE $\gamma$-rays between
    17 April to 1 June 2012 from HBL 1ES 1959+650 \citep{Aliu:2014hra}. On 20
    May, a short-lived VHE flare was detected which is fitted with the
    photohadronic model using $F_0=8.0\times 10^{-12} \, \rm{erg\,
      cm^{-2}\, s^{-1}}$, $\delta=2.9$. This corresponds to a high
    state and the intrinsic flux is shown in dashed line.}
  \label{fig:sfig2}
\end{figure}

\clearpage

\begin{figure}
  \centering
  \includegraphics[width=\linewidth]{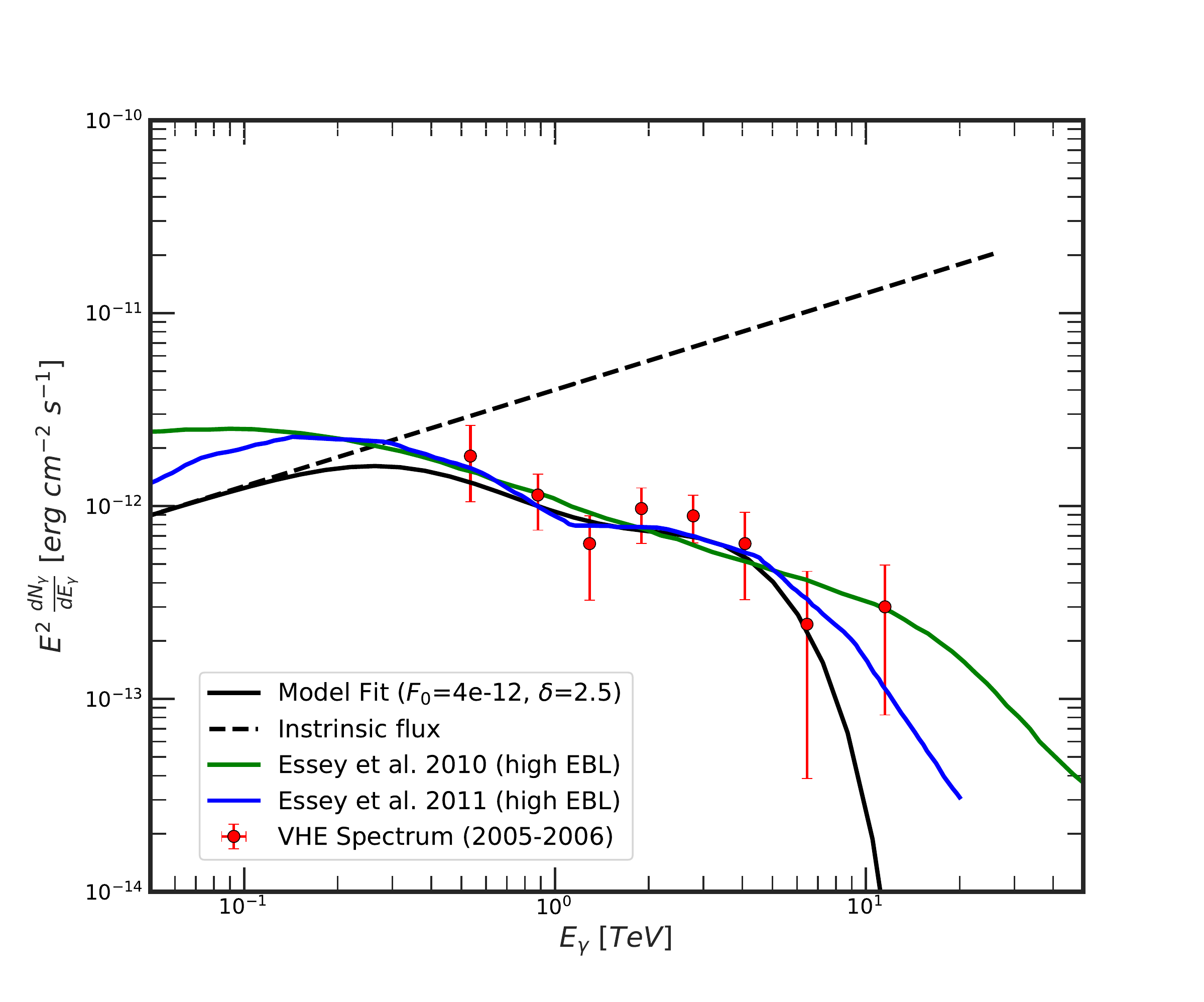}
  \caption{\small \textbf{Multi-TeV SED of 1ES 0229+200}.
    The HESS telescopes observed the HBL 1ES 0229+200 between
    2005 and 2006 for
    41.8 hours \citep{Aharonian:2017aa} and its VHE
    spectrum is shown, fitted with the
    photohadronic model with $F_0=4.0\times 10^{-12} \,
    \rm{erg\, cm^{-2}\, s^{-1}}$, $\delta=2.5$, corresponding to a very
    high emission state. Here we compare our results with the hadronic
    model of Essey et al. \citep{Essey:2009ju1,Essey:2010er1} which uses the EBL model of Stecker et
    al. (high EBL) \citep{Stecker:2006aj}. It is important to mention that the
    data points shown in \citep{Essey:2010er1} were
    slightly shifted to the left, so the model may not
    coincide exactly with the original data points as shown here.}
  \label{fig:sfig3}
\end{figure}

\clearpage

\begin{figure}
  \centering
  \includegraphics[width=\linewidth]{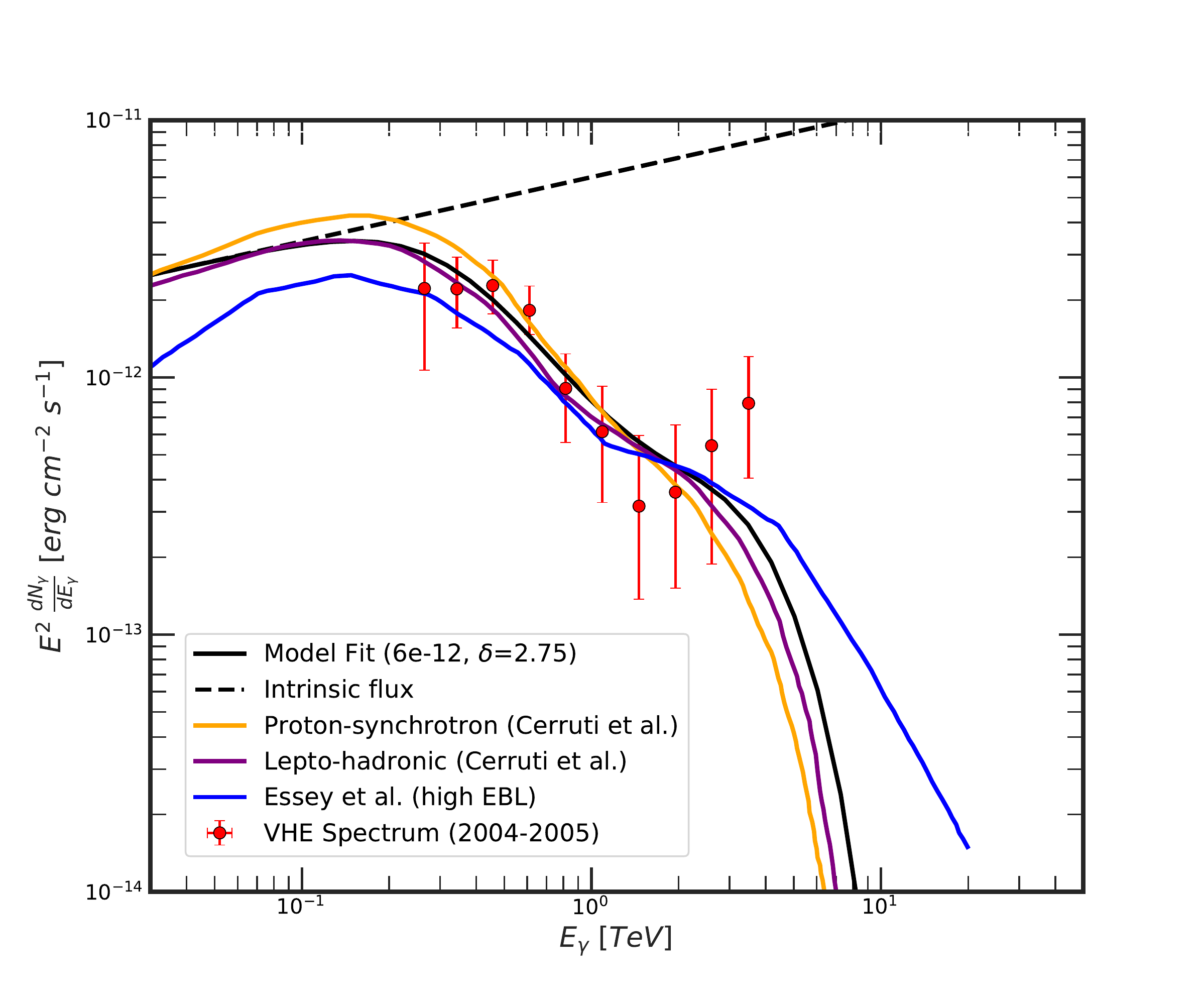}
  \caption{\small \textbf{Multi-TeV SED of 1ES 1101+232}.
    The HESS telescopes observed the HBL 1ES 1101+232 in 2004 for
    four nights in April for 2.7 hours, six nights in June
    for 8.4 hours and eleven nights in March 2005 for 31.6 hours \citep{Aharonian:2007nq}. The
    time-averaged VHE spectrum of these observations is fitted with the
    photohadronic model ($F_0=6.0\times 10^{-12} \, \rm{erg\, cm^{-2}\, s^{-1}}$,
    $\delta=2.75$) which corresponds to a high emission state and is compared with the hadronic model of Essey et
    al. \citep{Essey:2010er1} and the models of Cerruti et al. \citep{Cerruti:2014iwa}.}
  \label{fig:sfig4}
\end{figure}

\clearpage

\begin{figure}
  \centering
  \includegraphics[width=\linewidth]{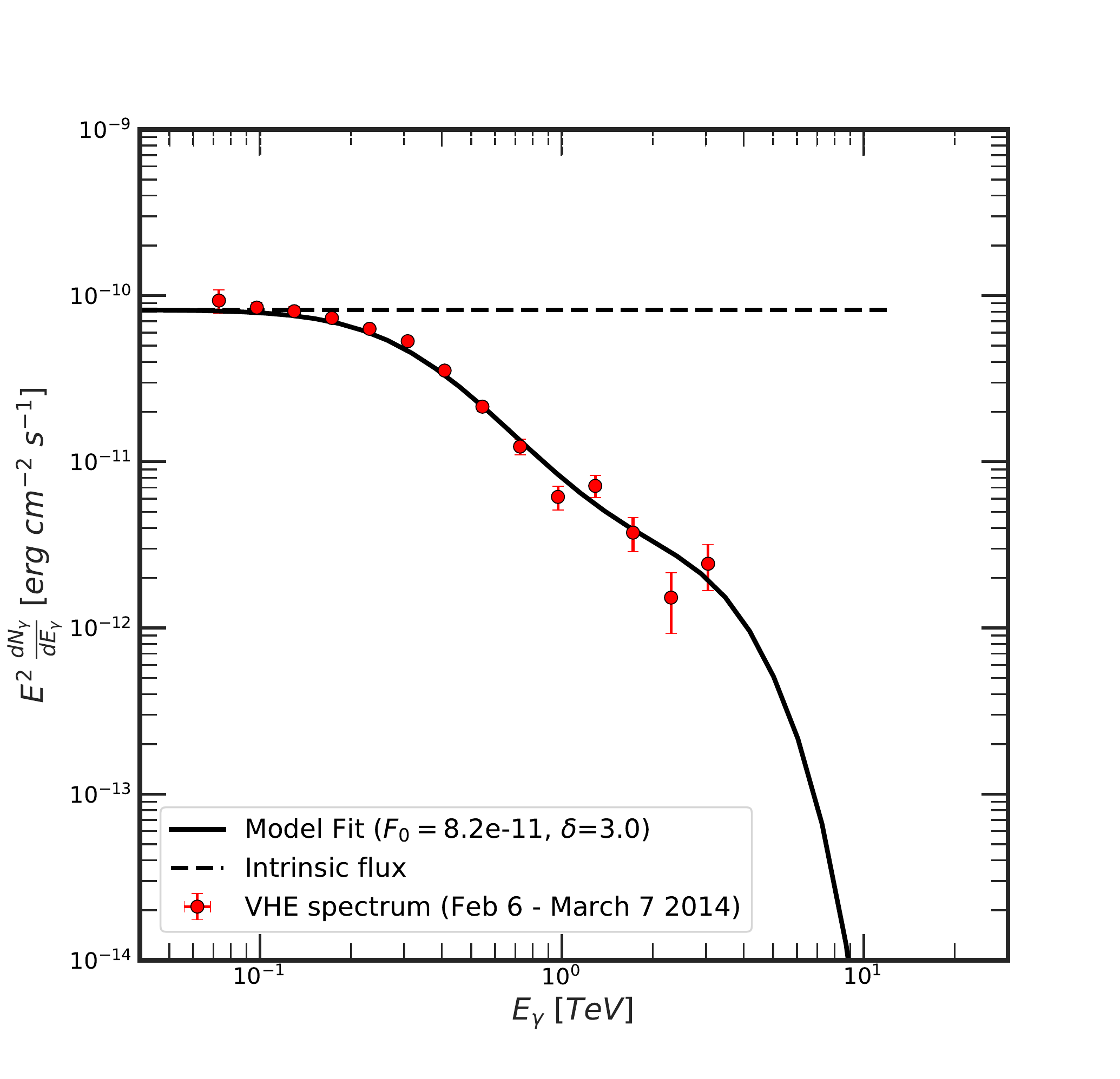}
  \caption{\small \textbf{Multi-TeV SED of 1ES 1011+496}.
    The HBL 1ES 1011+496 at a redshift of z=0.212 was observed by
    the MAGIC telescopes during a flaring event between February and
    March 2014, for a total of 17 nights \citep{Ahnen:2016gog}. In the photohadronic scenario a very good fit
    to the VHE spectrum is obtained for $F_0=8.2\times 10^{-11} \, \rm{erg\, cm^{-2}\,
      s^{-1}}$, $\delta=3.0$, which corresponds to a low state and thus a
    flat intrinsic spectrum.}
  \label{fig:sfig5}
\end{figure}

\clearpage
\begin{figure}
  \centering
  \includegraphics[width=\linewidth]{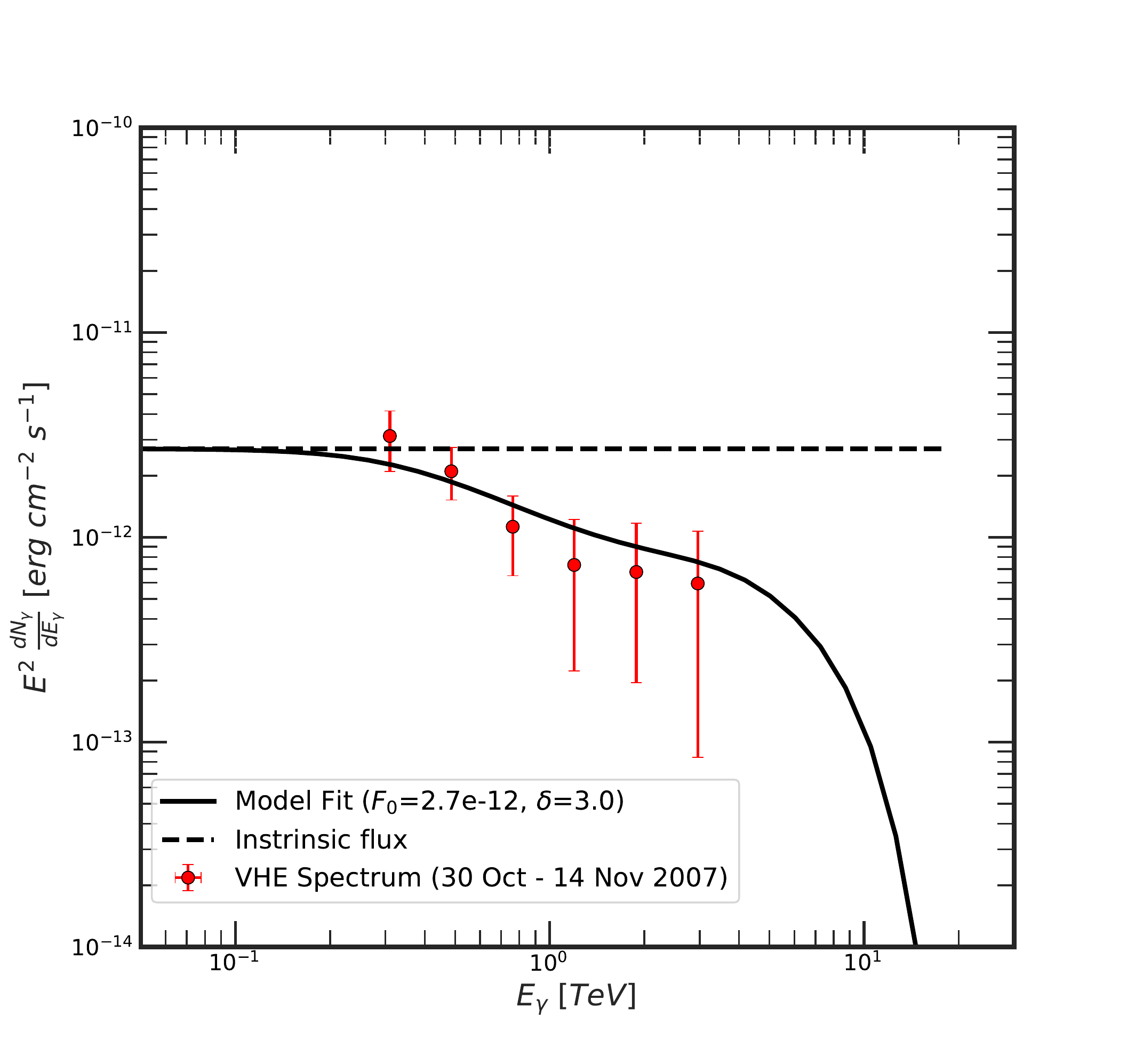}
  \caption{\small \textbf{Multi-TeV SED of RGB J0152+017}.
    The HBL RGB J0152+017 (z=0.8) is the farthest HBL in our
    list. The HESS telescopes observed during 30 October to 14
    November 2007 for a total of 14.7 hours and detected 173 VHE  $\gamma$-ray
    events \citep{Aharonian:2008aj}. We have shown the time-averaged VHE
    spectrum and fitted it with the photohadronic model for $F_0=2.7\times 10^{-12} \, \rm{erg\,
      cm^{-2}\, s^{-1}}$, $\delta=3.0$ which is a low emission state.}
  \label{fig:sfig6}
\end{figure}

\clearpage

\begin{figure}
  \centering
  \includegraphics[width=\linewidth]{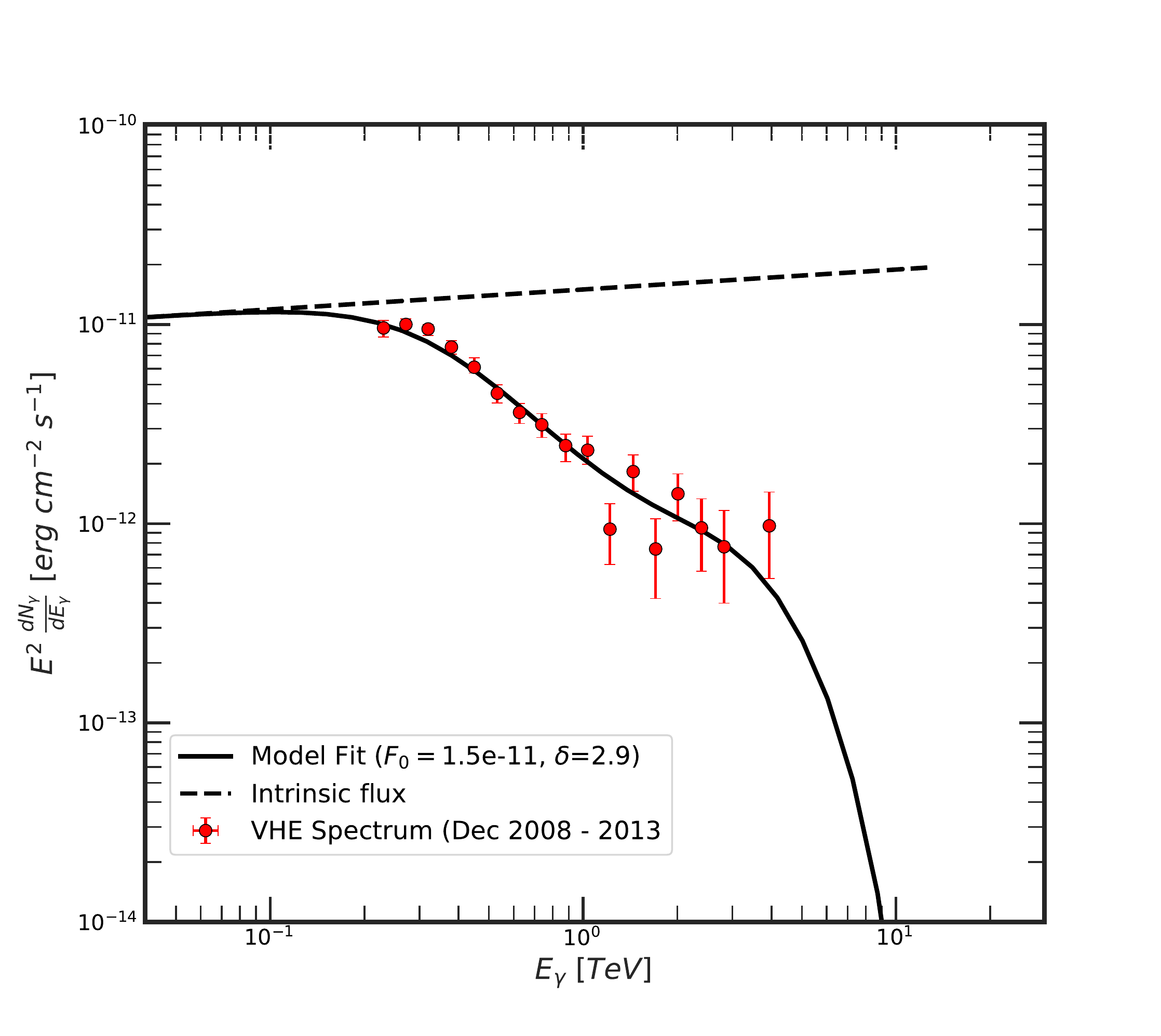}
  \caption{\small \textbf{Multi-TeV SED of 1ES 1218+304}.
    The HBL 1ES 1218+304 is at a redshift of z=0.182 and is a
    relatively bright source. It was observed by the VERITAS telescopes from December
    2008 until the 2012-2013 observing season \citep{Madhavan:2013sea},
    for a total of 86 hours. The time-averaged VHE
    spectrum is fitted well using the photohadronic model with
    $F_0=1.5\times 10^{-11} \, \rm{erg\, cm^{-2}\, s^{-1}}$,
    $\delta=2.9$, corresponding to a high state.}
  \label{fig:sfig7}
\end{figure}

\clearpage

\begin{figure}
  \centering
  \includegraphics[width=0.9\linewidth]{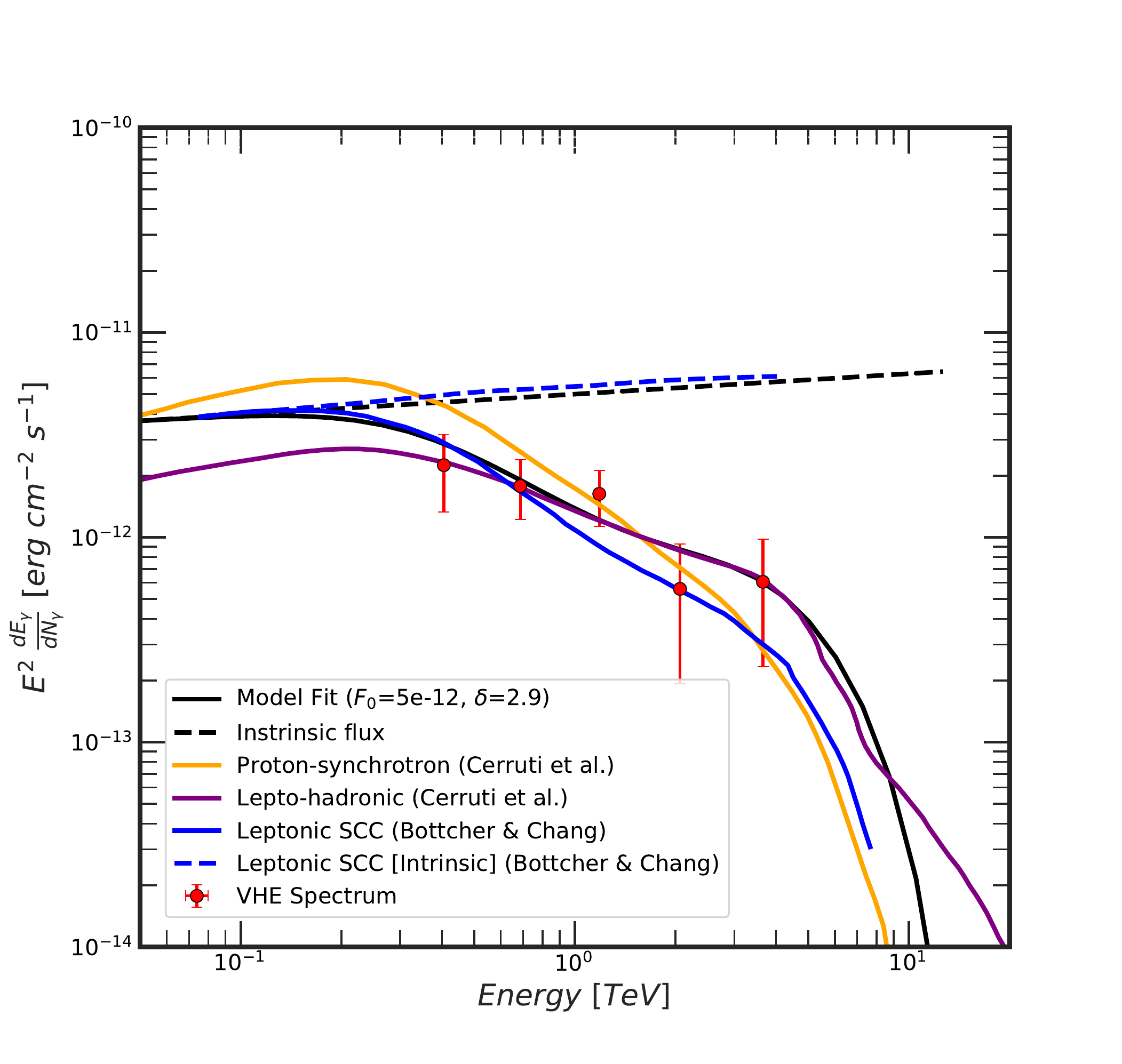}
  \caption{\small \textbf{Multi-TeV SED of RGB J0710+591}.
    The VERITAS telescopes observed RGB J0710+591 from December 2008 to March 2009
    for a total of 22.1 hours \citep{Acciari:2010qw}. The observed VHE
    spectrum is fitted by the proton-synchrotron and lepto-hadronic models
    of Cerruti {\it et al.} \citep{Cerruti:2014iwa} discussed in the main paper and the SSC model of
    Bötccher and Chang \citep{Acciari:2010qw}. Using our photohadronic model, we found an excellent fit to the data for
    $\delta=2.9$, which corresponds to a high emission state.}
  \label{fig:sfig8}
\end{figure}

\clearpage

\begin{figure}
  \centering
  \includegraphics[width=0.9\textwidth]{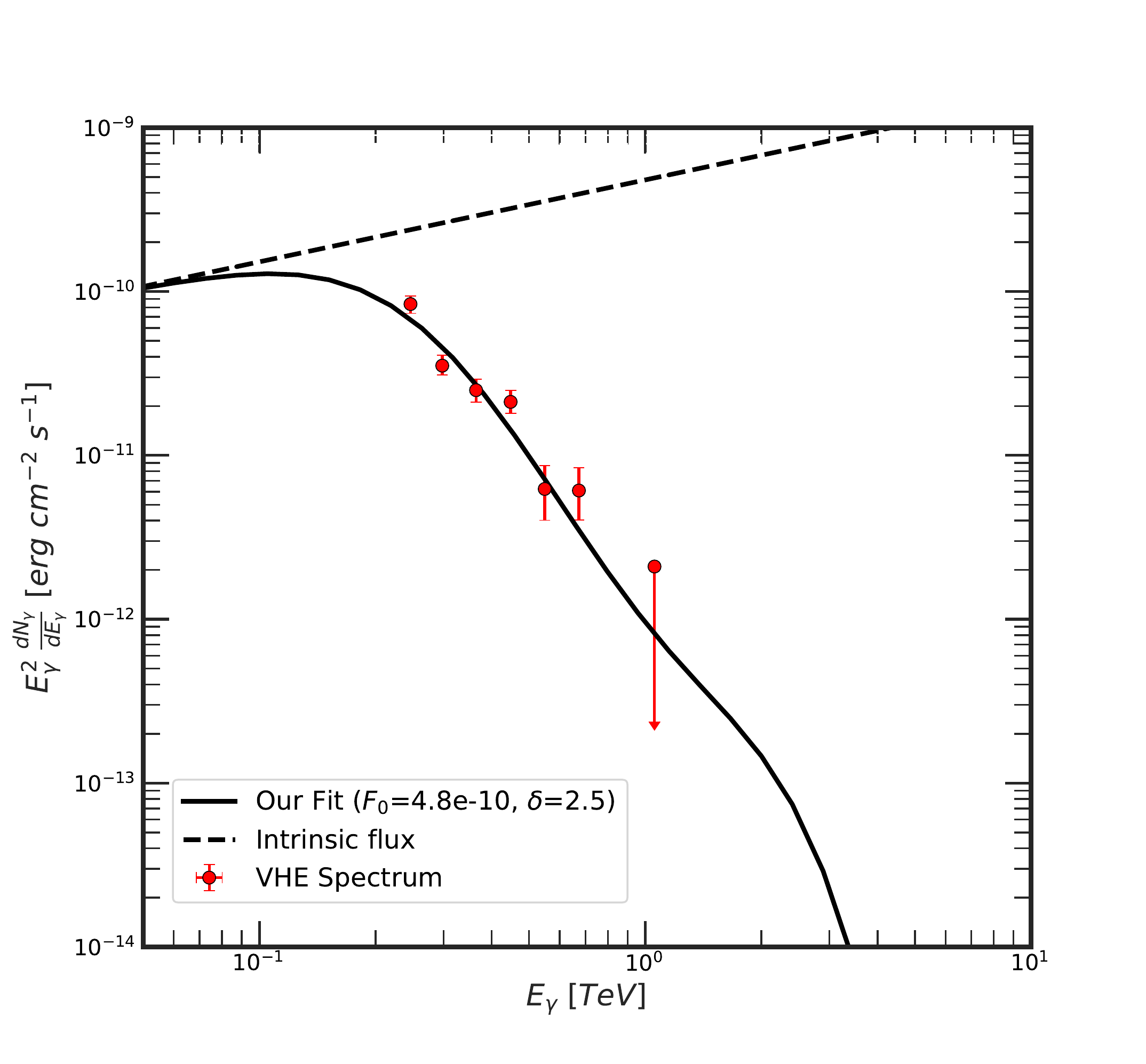}
  \caption{\small \textbf{Multi-TeV SED of PG 1553+113}.
    A multi-TeV flaring event was observed from PG 1553+113 during the nights of April 26 and 27 of 2012 by the HESS
    telescopes
    for a total of 3.5 hours in
    the energy range $0.25$ TeV to $0.67$ TeV
    \citep{Abramowski:2015ixa}. Its time-averaged photon flux 
    (red data points) is fitted well using the photohadronic model with 
    $F_0=4.8\times 10^{-10} \, \rm{erg\, cm^{-2}\, s^{-1}}$,
    $\delta=2.5$, corresponding to a very high state.}
  \label{fig:sfig9}
\end{figure}

\clearpage

\begin{figure}
  \centering
  \includegraphics[width=\linewidth]{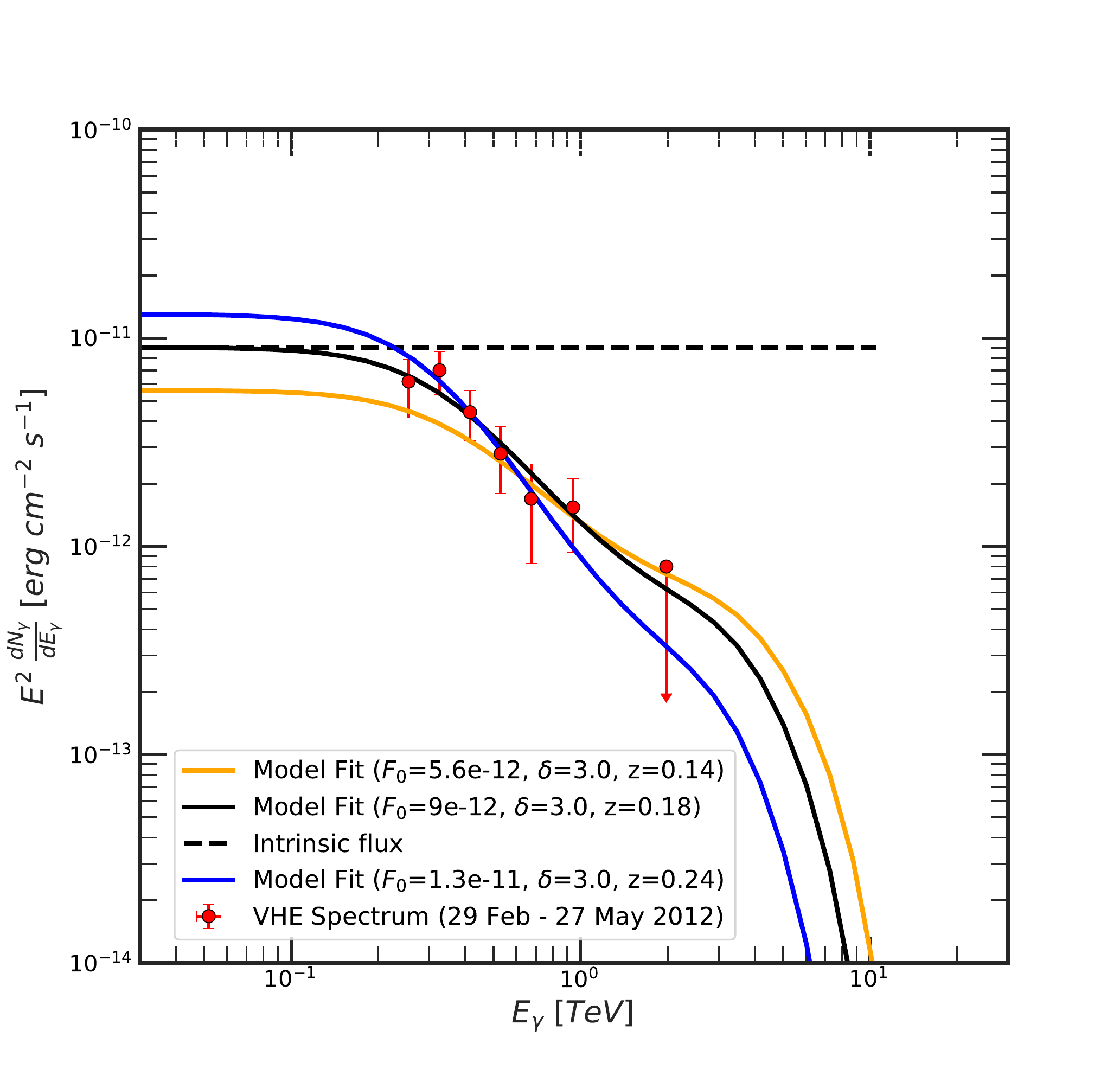}
  \caption{\small \textbf{Multi-TeV SED of PKS 1440-389}.
    The HBL PKS 1440-389 was observed by HESS telescopes between
    29 February to 27 May 2012 for a total of $\sim 12$ hours \citep{Prokoph:2015mza}. Due to
    poor spectral quality, the redshift of this object is not well known
    and the current best constraint is $0.14 < z <
    2.2$ \citep{Shaw:2013pp}. Using the photohadronic model
    and performing a statistical analysis for
    different redshifts, we constrained the 
    redshift in the range $0.14\le z\le 0.24$. We observed that for all
    these redshifts the value of $\delta=3.0$, which is a low emission
    state.}
  \label{fig:sfig10}
\end{figure}

\clearpage

\begin{figure}
  \centering
  \includegraphics[width=0.9\linewidth]{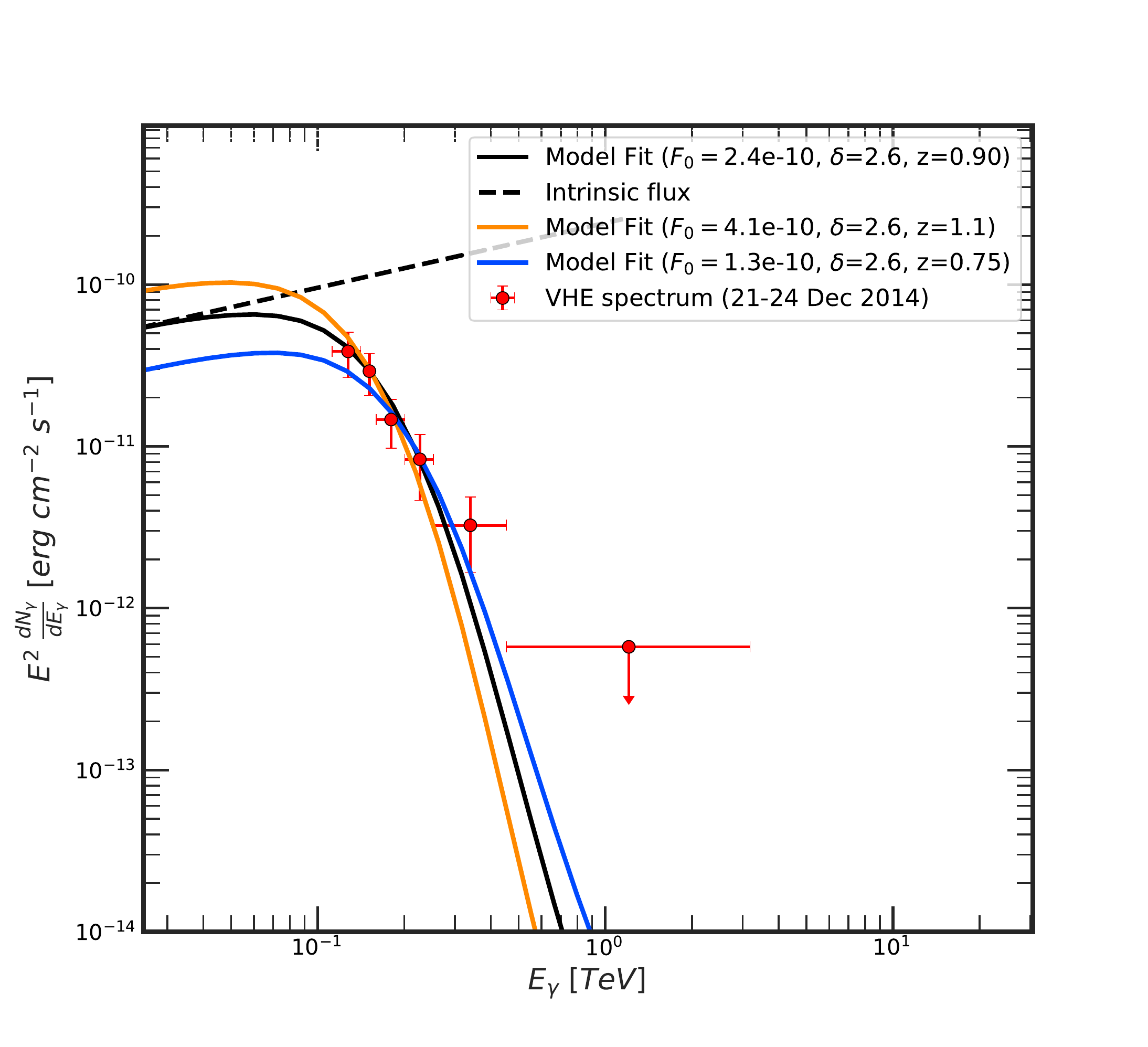}
  \caption{\small \textbf{Multi-TeV SED of RGB J2243+203}.
    The HBL RGB J2243+203 has an unknown redshift. Using different EBL models, the Fermi-LAT put an
    upper limit on the redshift ($z=1.1$). On 21 December 2014, the VERITAS
    telescopes observed elevated VHE flux and continued
    observing till 24 of December \citep{Abeysekara:2017moq}. 
    Using the
    photohadronic model and performing a statistical analysis for
    different cases, we were able to constraint the 
    redshift in the range $0.75\le z\le 1.1$. The best fit to the data
    was found for $z=0.90$ and $\delta=2.6$, which corresponds to a very high
    emission state.}
  \label{fig:sfig11}
\end{figure}


\clearpage

\begin{table}
  \label{table.1}
  \footnotesize
  \begin{center}
    \begin{tabular}{cccccc}
      Name & Redshift($z$) & Period & State & Ref. \\ [1 mm]
      \hline 
      Mrk 421 & 0.031 & 2004 & High & \citep{Blazejowski:2005ih} \\
      \ & \ & 22 Apr 2006 & High &  \citep{Aleksic:2010xj} \\ 
      \ & \ & 24 Apr 2006 & Low &  \citep{Aleksic:2010xj} \\ 
      \ & \ & 25 Apr 2006 & High &  \citep{Aleksic:2010xj} \\ 
      \ & \ & 26 Apr 2006 & Low &  \citep{Aleksic:2010xj} \\ 
      \ & \ & 27 Apr 2006 & High &  \citep{Aleksic:2010xj} \\
      \ & \ & 28 Apr 2006 & Low &  \citep{Aleksic:2010xj} \\
      \ & \ & 29 Apr 2006 & Low &  \citep{Aleksic:2010xj} \\ 
      \ & \ & 30 Apr 2006 & Very High &  \citep{Aleksic:2010xj} \\ 
      \ & \ & 16 Feb 2010 & Low &  \citep{Singh:2014yba} \\ 
      \ & \ & 17 Feb 2010 & Low &  \citep{Galante:crp} \\
      \ & \ & 10 Mar 2010 & Very High &  \citep{Aleksic:2014rca} \\
      \ & \ & 10 Mar 2010 &  Low &  \citep{Aleksic:2014rca} \\ 
      \ & \ & 28 Dec 2010 & Low &  \citep{Singh:2018tzy} \\

      Mrk 501 & 0.034 & 22 - 27 May 2012& High &  \citep{Chandra:2017vkw} \\ 

      1ES 2344+514 & 0.044 & 4 Oct 2007 - 11 Jan 2008 & Low &  \citep{Allen:2017cgm} \\

      1ES 1959+650 & 0.048 & May 2002 & Low &  \citep{Aharonian:2003be} \\  
      \ & \ & Nov 2007 - Oct 2013 & Low &  \citep{Aliu:2013aas} \\ 
      \ & \ & 21-27 May 2006 & Low &  \citep{Albert:2008uda} \\ 
      \ & \ & 20 May 2012 & High &  \citep{Aliu:2014hra} \\  

      1ES 1727+502 & 0.055 & 1-7 May 2013 & Low &  \citep{Archambault:2015sla} \\ 

      PKS 1440-389 & 0.14$\le$$z$$\le$0.24 & 29 Feb - 27 May 2012 & Low &  \citep{Prokoph:2015mza} \\ 

      1ES 1312-423 & 0.105 & Apr 2004 - Jul 2010 & Low &  \citep{Abrawoski:2013hss} \\ 

      B32247+381 & 0.119 & 30 Sep - 30 Oct 2011 & Low &  \citep{MAGIC:2012ac} \\ 

      RGB J0710+591 & 0.125 & Dec 208 - Mar 2009 & High &  \citep{Acciari:2010qw} \\ 

      1ES 1215+303 & 0.131 & Jan - Feb 2011 & Low &  \citep{Aleksic:2012npa} \\

      1RXS J101015.9-311909 & 0.14 & Aug 2008 - Jan 2011 & High &  \citep{Abrawoski:2012hss} \\ 

      1ES 0229+200 & 0.14 & 2005 - 2006 & Very High &  \citep{Aharonian:2017aa} \\

      H 2356-309 & 0.165 & Jun - Dec 2004 & High &  \citep{Aharonian:2005gh} \\ 

      1ES 1218+304 & 0.182 & Dec 2008 - 2013 & High &  \citep{Madhavan:2013sea} \\

      1ES 1101+232 & 0.186 & 2004 - 2005 & High &  \citep{Aharonian:2007nq} \\ 

      1ES 1011+496 & 0.212 & 6 Feb - 7 Mar 2014 & Low &  \citep{Ahnen:2016gog} \\ 

      1ES 0414+009  & 0.287 & Aug 2008 - Feb 2011 & High &  \citep{Madhavan:2013sea} \\ 

      PG 1553+113 & 0.50 & 26 - 27 Apr 2012 & Very high &  \citep{Abramowski:2015ixa} \\ 

      RGB J0152+017 & 0.80 & 30 Oct - 14 Nov 2007 & Low &  \citep{Aharonian:2008aj} \\

      RGB J2243+203 & 0.75$\le$$z$$\le$1.1 & 21-24 Dec 2014 & Very High &  \citep{Abeysekara:2017moq} \\ [1 mm]
      \hline
    \end{tabular}
    \caption{\small The flaring states of the HBLs given in Table 1 of the
      main article are given here along with their respective references.}
  \end{center}
\end{table}
\clearpage

\bibliography{Sup_v1.1}{}
\bibliographystyle{aasjournal}
